\documentclass{PoS}

\title{GPDs at HERA and perspectives at COMPASS}

\ShortTitle{GPDs at HERA and perspectives at COMPASS}

\author{\speaker{Laurent SCHOEFFEL}%
         \thanks{A footnote may follow.}\\
        CEA Saclay, Irfu/SPP, France \\
        E-mail: \email{laurent.schoeffel@cea.fr}}

\abstract{
Measurements of the deep-inelastic scattering (DIS) of leptons and nucleons, $e+p\to e+X$,
allow the extraction of Parton Distribution Functions (PDFs) which describe
the longitudinal momentum carried by the quarks, anti-quarks and gluons that
make up the fast-moving nucleons. 
While PDFs provide crucial input to
perturbative Quantum Chromodynamic (QCD) calculations of processes involving
hadrons, they do not provide a complete picture of the partonic structure of
nucleons. 
In particular, PDFs contain neither information on the
correlations between partons nor on their transverse motion.
Hard exclusive processes, in  which the
nucleon remains intact, have emerged in recent years as prime candidates to complement
this essentially one dimentional picture. 
The simplest exclusive process is the deeply virtual
Compton scattering (DVCS) or exclusive production of real photon, 
$e + p \rightarrow e + \gamma + p$.
This process is of particular interest as it has both a clear
experimental signature and is calculable in perturbative QCD. 
The DVCS reaction can be regarded as the elastic scattering of the
virtual photon off the proton via a colourless exchange, producing a 
real photon in the final state  \cite{dvcsh1,dvcszeus}. 
In the Bjorken scaling 
regime, 
QCD calculations assume that the exchange involves two partons, having
different longitudinal and transverse momenta, in a colourless
configuration. These unequal momenta or skewing are a consequence of the mass
difference between the incoming virtual photon and the outgoing real
photon. This skewedness effect can
 be interpreted in the context of generalised
parton distributions (GPDs) \cite{qcd}. 
In this proceeding, 
we examine  typical measurements from HERA and prospects for
COMPASS at CERN \cite{dhose}, that can bring new insights
on the quarks/gluons imaging of the nucleon.
}

\FullConference{European Physical Society Europhysics Conference on High Energy Physics\\
                 July 16-22, 2009\\
                 Krakow, Poland}

\begin{document}

\section{Introduction}

\begin{figure}[!htbp] 
  \begin{center}
    \includegraphics[width=7cm]{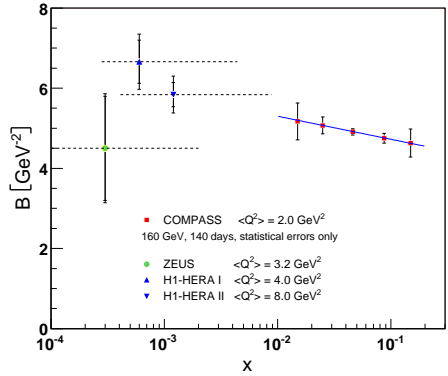}
  \end{center}
  \caption{The logarithmic slope of the $t$ dependence
  for DVCS exclusive production, $b$ as a function of $x_{Bj}$, 
  extracted from a fit
  $d\sigma/dt \propto \exp(-b|t|)$  where $t=(p-p')^2$.
  H1 and ZEUS points are measurements and simulations are
  displayed for COMPASS (CERN).
}
\label{fig2b}  
\end{figure}

A major experimental achievement 
of H1 and ZEUS \cite{dvcsh1,dvcszeus} has been the measurement of
DVCS cross sections, differential in $t=(p'-p)^2$, 
the momentum transfer (squared) at the proton vertex.
A good description
of $d\sigma_{DVCS}/dt$ by a fit of the form $e^{-b|t|}$
is obtained \cite{dvcsh1,dvcszeus}. 
Hence, an extraction of the $t$-slope parameter $b$ is accessible
and it can be achieved experimentally. Measurements at HERA
and simulations  for the COMPASS kinematics are displayed in Fig. \ref{fig2b}
 \cite{dhose}.
We observe the good complementarity of both kinematical coverage in $x_{Bj}$.

\section{On the interest of $t$ slope measurement}

Measurements of the $t$-slope parameters $b$
are key measurements for almost all exclusive processes,
in particular DVCS.
 Indeed,
a Fourier transform from momentum
to impact parameter space readily shows that the $t$-slope $b$ is related to the
typical transverse distance between the colliding objects \cite{qcd}.
At high scale, the $q\bar{q}$ dipole is almost
point-like, and the $t$ dependence of the cross section is given by the transverse extension 
of the gluons (or sea quarks) in the  proton for a given $x_{Bj}$ range.
More precisely, from GPDs, we can compute
a parton density which also depends on a spatial degree of freedom, the transverse size (or impact parameter), labeled $R_\perp$,
in the proton. Both functions are related by a Fourier transform 
$$
PDF (x, R_\perp; Q^2) 
\;\; \equiv \;\; \int \frac{d^2 \Delta_\perp}{(2 \pi)^2}
\; e^{i ({\Delta}_\perp {R_\perp})}
\; GPD (x, t = -{\Delta}_\perp^2; Q^2).
$$
Thus, the transverse extension $\langle r_T^2 \rangle$
 of gluons (or sea quarks) in the proton can be written as
$$
\langle r_T^2 \rangle
\;\; \equiv \;\; \frac{\int d^2 R_\perp \; PDF(x, R_\perp) \; R_\perp^2}
{\int d^2 R_\perp \; PDF(x, R_\perp)} 
\;\; = \;\; 4 \; \frac{\partial}{\partial t}
\left[ \frac{GPD (x, t)}{GPD (x, 0)} \right]_{t = 0} = 2 b
$$
where $b$ is the exponential $t$-slope.
Measurements of  $b$
presented in Fig. \ref{fig2b} for HERA
corresponds to $\sqrt{r_T^2} = 0.65 \pm 0.02$~fm at large scale 
$Q^2$ for $x_{Bj} < 10^{-2}$.
This value is smaller that the size of a single proton, and, 
in contrast to hadron-hadron scattering, it does not expand as energy $W$ increases.
Obviously, it would be interesting to perform similar measurements
in the COMPASS kinematic domain to obtain the information at intermediate $x_{Bj}$
 \cite{dhose}.

Let us note that  HERA results are consistent with perturbative QCD calculations in terms 
of a radiation cloud of gluons and quarks
emitted around the incoming virtual photon.
The fact the perturbative QCD calculations provide correct descriptions
of $b$ measurements is a proof that
they deal correctly this this non-trivial aspect of the proton 
(spatial) structure. 
The modeling of the correlation between the spatial transverse
structure and the longitudinal momenta distributions of partons in the proton
is one major challenge  for the GPDs approach. 

Another natural  way to address the problem 
of the correlation  between $x$ and $t$ kinematical variables
proceeds from
a determination of a cross section asymmetry with respect to the beam
charge. It has been performed recently by the H1 experiment by measuring the ratio
$A_C=(d\sigma^+ -d\sigma^-)/ (d\sigma^+ + d\sigma^-)$ as a function of $\phi$,
where $\phi$ is the azimuthal angle between leptons and proton plane.
The result is presented in Ref. \cite{dvcsh1}.
$A_C$ is found to be
$$
A_C \simeq 0.16 \cos \phi
$$
at low $x_{Bj}<0.01$.
This result represents  a major experimental progress 
and is challenging for models. 
Let us note that models of GPDs can use
 present HERA data at low $x_{Bj}$,
as well as JLab and HERMES data at larger $x_{Bj}$ ($x_{Bj}>0.1$),
in order to provide a first global understanding of exclusive real photon
production \cite{qcd}.  
However, as already mentioned above, some efforts have
still to be made in the intermediate $x_{Bj}$ domain  \cite{dhose}.

\section{Summary and outlook}
DVCS measurements in the HERA kinematics at low $x_{Bj}$ ($x_{Bj}<0.01$)
are well described by recent GPDs models, which also describe
correctly measurements at larger values of $x_{Bj}$ in
the JLab kinematics.
Recently, H1 and ZEUS 
experiments have also shown that proton 
tomography at low $x_{Bj}$
enters into the experimental domain of high energy physics, with a first 
experimental evidence
that gluons are located at the periphery of the proton. A new frontier in 
understanding
this structure would be possible at CERN within the 
COMPASS experimental setup. Major advances have already been done
on the design of the project and simulation outputs.

\end{document}